\documentclass[aps,prx,reprint,superscriptaddress]{revtex4-1}
\usepackage[pdftex]{hyperref,graphicx}	% using pdflatex
\usepackage{dcolumn}   	\usepackage[pdftex]{hyperref,graphicx}	% using pdflatex
\usepackage{bm}
\usepackage{natbib}
\usepackage{array}
\usepackage[active]{srcltx}
\usepackage{subfigure,amssymb,amsmath,natbib, mathtools} %,floatflt}
\usepackage{graphicx,varioref}
\usepackage{epstopdf}
\usepackage{color}
\usepackage{textcomp} %for use of \textmu
\usepackage[utf8]{inputenc}

\newcommand{\bse}{\begin{subequations}} 
\newcommand{\ese}{\end{subequations}} 
\newcommand{\be}{\begin{equation}} 
\newcommand{\ee}{\end{equation}} 
\newcommand{\bea}{\begin{eqnarray}} 
\newcommand{\eea}{\end{eqnarray}}

\begin{document}
\title{Viscous energy dissipation in slender channels with porous or semipermeable walls}
\author{Hanna Rademaker, Kaare H. Jensen and Tomas Bohr}
\affiliation{Department of Physics, Technical University of Denmark, 2800 Kgs. Lyngby, Denmark\\ \today}

\begin{abstract}
\begin{center}

\end{center}
We study the viscous dissipation in pipe flows in long channels with porous or semipermeable walls, taking into account both the dissipation in the bulk of the channel and in the pores. We give simple closed form expressions for the dissipation in terms of the axially varying flow rate $Q(x)$ and the pressure $p(x)$, generalizing the well known expression $\dot W=Q\,\Delta p$ for the case of impenetrable walls with constant $Q$ and a pressure difference $\Delta p$ between the ends of the pipe. When the pressure $p_0$ outside the pipe is constant, the result is the straightforward generalization $\dot W=\Delta \left[(p-p_0) \,Q\right]$.  
Finally, applications to osmotic flows are considered.
\end{abstract}

%\begin{keywords}
%Osmotic flow. Stagnant zone. Flows in pine needles. Dissipation in porous pipe. 
%\end{keywords}
\maketitle

\section{Introduction}
Channel flows -- liquid flows confined within a closed conduit with no free surfaces -- are omnipresent. In animals \citep{LaBarbera1990} and plants \citep{Holbrook2005} they serve as the building blocks of vascular systems, distributing energy to where it is needed and allowing distal parts of the organism to communicate.
When constructed by humans, one of the major functions of channels is to transport liquids or gasses,  e.g. water (irrigation and urban water systems) and energy (oil or natural gas) from sites of production to the consumer or industry.

In some cases, the channels have solid walls which are impermeable to the liquid flowing inside. In other cases, the channels have porous walls which allow the liquid to move across the wall and thus modify the axial flow. If solutes are present in the liquid, the walls can be semipermeable, allowing only the solvent to pass and thereby allow filtration or create osmotically driven flows due to concentration differences between the inside and the outside.  Flows with impermeable walls have been studied in great detail, and analytical solutions are known in a few, but important, cases \citep{Batchelor2000}. Flows with porous walls have received much less attention, although they are equally important. The effect of porous walls is especially important in the study of biological flows  \citep{Holbrook2005, Marbach2016, RMP2016}  and in industrial filtration applications \citep{Nielsen2012}.

Exact solutions for the flow in porous walled channels are known in a few important cases. Berman's method \cite{Berman1953} allows for the solution of steady flows in geometries with symmetries, for example, between parallel plates or in a cylindrical tube. The technique is closely related to those commonly used in boundary layer theory \citep{Schlichting2000}. By demanding that the solution be of similarity form, Berman's method reduces the Navier-Stokes to a single non-linear third-order differential equation for the velocity potential in one space dimension. The flow between parallel plates \citep{Berman1953} and in a cylindrical \citep{Yuan1956a,Aldis1988} and annular tube \citep{Berman1958} have been analyzed in this way. 
Time dependent flows,  high-Reynolds-number flows, and stability and uniqueness of the solutions have since been address by a large number of workers using analytical and numerical methods, see e.g. \citet{Cox1991,King2001,Majdalani2003,Dauenhauer2003,Kurdyumov2008,Saad2009,Xu2010,Liu2011}. 

Despite our broad knowledge of transport characteristics in porous channel flows, the energetic cost of flow remains poorly understood. In conventional low-Reynolds-number pipe flows, the link between the flow rate $Q$, pressure drop $\Delta p$ and energy dissipation rate is $\dot W=Q\Delta p$, analogous to an electrical circuit. However, in porous channels, both the flow rate and pressure are position-dependent, hence the standard result is inadequate. 

In this paper we shall concentrate on the case of a long cylindrical pipe or tube with porous or semipermeable walls. We first (Section II) discuss the basic fluid dynamics based on the solution by Aldis \cite{Aldis1988} for a long cylindrical porous pipe. In Section III, we write down the general expression for the viscous dissipation both in the bulk of the pipe and in its porous walls. Finally, in the last section, we discuss two specific examples, one where the porous inflow is constant, and and one where the external conditions are constant.

\section{Low-Reynolds-number flow in a long cylindrical, porous pipe}
%{\color{red} Maybe give NVS first here, then proceed to aldis flow}

\begin{figure}
  \centering
  \includegraphics[width=\columnwidth]{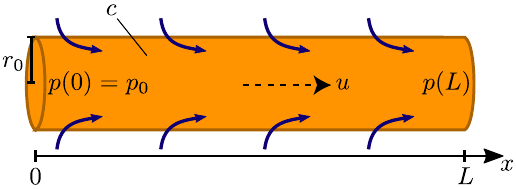} 
  \caption{A tube of length \(L\) and circular cross-section of radius \(r_0\), is filled with a liquid of viscosity $\eta$ and density $\rho$. Liquid is injected from the surrounding medium (arrows) and creates a bulk flow of speed $u$ along the positive $x$-direction of the tube. }
  \label{fig:singleTube}
\end{figure}

We consider a tube of length $L$ and characteristic transverse dimension $r_0$ embedded in a fluid-saturated medium (Fig.~\ref{fig:singleTube}). 
The channel walls of thickness $d$ are permeable, characterized by the Darcy permeability $k$, such that the trans-membrane velocity field is normal to the channel walls
\begin{align}
\nonumber
    {\mathbf v}_m &= - \frac{k}{\eta d}\left(p_e(x)-p(x)\right) {\hat{\mathbf n}} \\
    \label{vm1}
    &= - L_p\left(p_e(x)-p(x)\right) {\hat{\mathbf n}}  =  L_p\, \delta p(x) \, {\hat{\mathbf n}}  
\end{align}
where $p_e$ and $p(x)$ is the external medium and channel pressure, respectively, $\delta p = p - p_e$ is the pressure drop across the membrane, and ${\hat{\mathbf n}}$ is the outward normal (i.e., $\hat{{\mathbf r}}$ in cylindrical coordinates) Further, $L_p = k/(\eta d)$ is the {\em membrane permeability}. A detailed model of $L_p$, assuming parallel cylindrical pores, is given in Appendix A.  

When we are dealing semipermeable membranes separating solutions with different solute concentration, (\ref{vm1}) takes the form
\begin{equation}
\label{vm2}
     {\mathbf v}_m  = - L_p\left(\Psi_e(x)-\Psi(x)\right) \mathbf n = L_p\, \delta \Psi(x)
\end{equation}
where  $\Psi $ is the {\em water potential}. The water potential (free energy) $\Psi = p - \Pi $ includes the osmotic pressure $\Pi$, which, for low solute concentrations $c$, can be expressed using the van't Hoff relation $\Pi = R T c$ (see e.g,. \cite{Fermi1956}). Again, $\delta$ denotes the jump accross the membrane: $\delta \Psi = \Psi(x)-\Psi_e(x)$.
In the rest of the paper, we assume separation of geometric scales, such that the channel is long in comparison to all other lengths, i.e., $L\gg r_0 \gg d \gg \sqrt k$.

In steady low-Reynolds-number flow conditions, we base our analysis on the Stokes equation
\be
\nabla p = \eta \nabla^2 {\bf v},
\ee
for an incompressible fluid where $ \nabla \cdot {\bf v} = 0$ and a constant viscosity $\eta$ and density $\rho$. Note that the no-slip conditions on the channel boundaries correspond to $\mathbf v_\perp = \mathbf v_m $ and $\mathbf v_\parallel = \mathbf 0$.

We express the velocity using axial and radial coordinates, i.e., $\mathbf v = (v_x,v_r)$ and thus assume rotational symmetry.  In plants, the sieve tubes of the phloem are roughly of this form, and in the leaves their radii ($r_0$) are in the $\mu$m regime while their length ($L$) is centimetric. The slender (lubrication) approximation used by \citet{Aldis1988}  to describe such flows is valid, since $Re\ll 1$ and $r_0/L \ll 1$. The 
stationary flow field then has the form 
\begin{align}
\label{AldisA}
 v_r(r,x) &= f ( r) v_0(x) \\
 \label{AldisB}
 v_x(r,x) &=g( r) u(x)
\end{align}
where 
\begin{align}
\label{Aldisf}
 f( r) &= \frac{r^3}{r_0^3} - 2\frac{r}{r_0}\\
 \label{Aldisg}
g( r) &= 2 \left( 1 - \frac{r^2}{r_0^2} \right)  
\end{align}
The function $v_0(x)$ is the radial injection velocity
 \be
v_r(r_0,x) = -v_0(x) = v_m(x)
 \label{Osmoin}
 \ee
 where $v_m$ given by (\ref{vm1}) or (\ref{vm2}) such that a positive $v_0$ denotes an {\em inflow}.
The mean axial flow speed is
 \be
\label{av}
u(x)  = \frac{2\pi}{\pi r_0^2}\int_0^{r_0}v_x(r,x)\, r\mathrm dr= {\bar v_x(r,x)}
\ee
and the corresponding volumetric flow rate $Q$ is
 \be
 Q(x)=\pi r_0^2 u(x) = Q_0+2 \pi r_0 \int_0^x v_0(x') dx'
 \ee
 where $Q_0$ is the inlet flow rate at $x=0$. The average flow speed is
  \be
\label{av1}
u(x) = \frac{Q_0}{\pi r_0^2}+ \frac{2}{r_0} \int_0^x v_0(x') dx' 
\ee
or
\begin{align}
\label{A3}
u'(x) = \frac{2}{r_0}v_0(x)
\end{align}
and 
\begin{align}
\label{AldisQ}
Q'(x) = 2 \pi r_0 v_0(x).
\end{align}
In the lubrication approximation the pressure does not vary over the pipe cross-section, i.e. we can replace $p$ by it's average value $p(x) = {\bar p(r,x)}$.
Thereby the average velocity is related to the axial pressure gradient as in standard Hagen-Poiseuille flow
\be
\label{Darcy1}
\frac{dp}{dx}= -\frac{8 \eta}{r_0^2} u(x) = - \frac{8 \eta}{\pi r_0^4} Q(x)
\ee
and, finally, the inflow $v_0 (x) =- v_m$ is given by (\ref{vm1}) or (\ref{vm2}) depending on whether the membrane is fully permeably or only permeable to the solvent.

\section{Viscous dissipation}
We shall determine the viscous dissipation in the flows studied in Section II, by looking firstly at the dissipation in the bulk flow, and secondly at the flow through the porous semipermeable walls, and then putting them together. Finally, we shall verify these expressions, by looking at the energy advection equation. The dissipated energy is given as
(see e.g., \cite{LL1987})
\be
\label{energy1}
 \dot{W} = \frac12 \eta \int \sum_{i,j} u_{ij}^2 dV =\frac12 \eta \int Tr \left[ {\rm \bf{u}} \cdot {\rm \bf{u}}^{T} \right] dV 
\ee
where $u_{ij}$ is the strain rate
\be
\label{strainrate}
u_{ij} = \left(\frac{\partial v_i}{\partial x_j} + \frac{\partial v_j}{\partial x_i}\right)
\ee
and where the volume integral goes over the volume of the flow. The sum in (\ref{energy1}), being the trace of the product of u-matrices, is invariant with respect to transformations to other (locally) orthogonal coordinates, and in particular, $i$ and $j$ can represent the cylindrical coordinates used above.

Note that this dissipative energy only represents the work done by the viscous forces in the fluid. For osmotically driven flows in plant leaves there would be an energy consumption related to the transport of sugar into the tubes, which we are not trying to account for here. 

The dissipation has the form
\begin{align}
\dot{W}_{\text{tot}}= \dot{W}_{\text{bulk}} + \dot{W}_{\text{walls}}
\end{align}
with terms coming coming from the bulk flow and from the flow in the porous walls.

The viscous dissipation for an axially symmetric flow, such as the Aldis flow field given by Eq. (\ref{AldisA})-(\ref{AldisB}),
can then be written as
\begin{align}
\nonumber
 \dot{W} &= 2\eta \int dV \left[ \left( \frac{\partial v_r}{\partial r} \right)^2 + \left( \frac{v_r}{r} \right)^2 \right.\\
 \label{dissip1}
 &+ \left.\left( \frac{\partial v_x}{\partial x} \right)^2 + \frac{1}{2} \left( \frac{\partial v_r}{\partial x} + \frac{\partial v_x}{\partial r} \right)^2 \right].
\end{align}
For the Aldis flow we can write the velocity components explicitly using (\ref{AldisA})-(\ref{AldisQ}). To obtain
this solution, we made the assumption that $v_r \ll v_x$ and $\partial/\partial x \ll \partial/\partial r$, so the dominant term in the dissipation is
\begin{align}
\nonumber
 \dot{W}_{\text{bulk}} = \eta \int dV   \left(  \frac{\partial v_x}{\partial r} \right)^2  &=  \eta \int dV   \left( g'( r) \right)^2  u^2(x)\\
 \label{dissip2}
  &= \frac{8 \eta}{\pi r_0^4}\int_0^L Q^2(x) dx
\end{align}
where we have used that $g'( r) = -4 r/r_0^2$ (\ref{Aldisg}). Using the Hagen-Poiseuille relation (\ref{Darcy1}) this can be written
\begin{align}
\label{dislub}
 \dot{W}_{\text{bulk}} &= -\int_0^L p'(x) Q(x) dx
\end{align}
and for a normal Poiseuille flow in a cylindrical pipe with solid walls (and therefore $Q$ and $p'=\Delta p/L$  constant) this becomes $  \dot{W}_{\text{bulk}}= Q \Delta p$ as it should.
The additional terms in (\ref{dissip1}) can be written in descending orders of $\left(L/r_0\right)^2$ as
 \begin{align}
 \nonumber
&\Delta \dot{W}_{\text{add}} =  \\
\nonumber
&\frac{1}{3\pi }\frac{\eta}{r_0^2} \left[ 5 \int_0^L \left(Q'\right)^2 dx + 8 \left(Q'(L)Q(L) - Q'(0)Q(0)\right) \right] \\
 & + \frac{11}{48 \pi}\eta \int_0^L \left(Q''\right)^2 dx 
 \label{eq:Wdot03c}
\end{align}
and in order of magnitude they correspond to replacing 2 or 4 factors of $r_0$ by factors of $L$ and it would thus not be justified to keep them in the lubrication limit used to obtain Eq. (\ref{AldisA})-(\ref{AldisB}).

To discribe the dissipation in the porous tube wall, we use Darcy's law (\ref{vm1}) or (\ref{vm2}) in the form
\begin{align}
\label{Darcy}
v_0(x) &= \frac{k}{\eta d} \delta p  =  L_p \delta p 
\end{align}
where $\delta p$ is the pressure jump across the porous tube wall, which, in osmotic flows, should be replaced by the jump in water potential $\delta \Psi = p-RT c$. The corresponding energy dissipation pr. unit wall area is simply $ \dot{w}_{\text{mem}} = v_0 \Delta p = (v_0(x))^2/L_p$
and the total dissipation is
\begin{align}
\nonumber
 &\dot{W}_{\text{wall}}=2 \pi r_0\int_0^L v_0(x)  \,\delta  p \, d x= \frac{2 \pi r_0}{ L_p}\int_0^L (v_0(x))^2 d x\\ 
 \label{dissipMem}
 &= \frac{1}{2 \pi r_0 L_p}\int_0^L (Q'(x))^2 d x = \int_0^L \delta p(x) \,Q'(x) d x
\end{align}

The total dissipation is found by adding the wall contribution (\ref{dissipMem}) to the bulk contribution (\ref{dissip2}) giving
\begin{align}
\label{genWp}
 \dot{W}_{\text{tot}}  &= -\int_0^L \left(p'(x) \, Q(x)  +  \delta  p(x) \, Q'(x)   \right)dx
\end{align}
which can also be written
\begin{align}
  \label{DissTot}
 \dot{W}_{\text{tot}}  & = \frac{8 \eta}{\pi r_0^4} \left( \int_0^L \left( Q^2(x) + L^2_{0} \left( Q'(x) \right)^2\right) \right)
\end{align}
where $L_{0}$ is the ``efficient length" introduced by Rademaker et al. in the context of osmotically driven pipe flows  \cite{Rademaker2017}, and, earlier, by Landsberg and Fowkes  in the context of water motion through root hairs  \cite{Landsberg1978}.
\be
\label{L0}
L_{0} = \left(\frac{ r_0^3}{16 \eta L_p} \right)^{1/2} = \left(\frac{ r_0^3 d}{16 k} \right)^{1/2} 
\ee

If we go back to the variables $p$ (or $\Psi$) and $Q$, we can write these expressions in a more general way. In order to treat the porous and the semipermeable case together, we shall write the inflow condition (\ref{vm1})-(\ref{vm2}) in general as
\begin{equation}
v_0(x) = - L_p \, \delta \Psi (x)
\end{equation}
where we, in the porous case simply assume that $c=0$. Then, using (\ref{AldisQ}) and (\ref{Darcy1}), we get
\begin{align}
\label{genW}
 \dot{W}_{\text{tot}}  &= -\int_0^L \left(p'(x) \, Q(x)  +  \delta \Psi (x) \, Q'(x)   \right)dx
\end{align}
an expression which, like (\ref{genWp}), is completely free of material parameters.

\section{Special cases}
\subsection{Constant inflow}
If we assume a constant inflow $v_0$, we have $Q(x) = Q_0 + 2 \pi r_0 v_0 x$ and 
\begin{align}
\nonumber
& \dot{W}_{\text{tot}}   = \frac{8 \eta}{\pi r_0^4} \left( \int_0^L \left( Q^2(x) + L^2_{0} \left( Q'(x) \right)^2\right) \right)\\
  \label{ConstQ}
&=32 \pi \frac{\eta v_0^2 L_0^3}{r_0^2} \left( \left((1+b^2) m + b  m^2 + \frac13  m^3  \right)\right)
\end{align}
where $m= L/L_{0}$  and $b=Q_0/(2 \pi r_0 v_0 L_{0})$ are dimensionless numbers.
For a tube closed in one end (like a pine needle) $Q_0=b=0$ and one can see that the membrane dissipation dominates for small $L \ll L_0$ and the
bulk dissipation dominates for large $L \gg L_0$.

\subsection{Constant external conditions}
If the pressure outside the tube $p_e$  is constant we can introduce the new pressure $p \to p-p_e$ in
(\ref{genWp})) and get simply
\begin{align}
\label{p-const}
 \dot{W}_\text{tot} =  \Delta[p\, Q] = p(0) Q(0) - p(L) Q(L)
\end{align}
generalizing the Hagen-Poiseuille result $ \dot{W} = Q\, \Delta p$.

Similarly, for the osmotic case, if both the pressure and concentration outside the tube $c_e$  are constant, we can introduce
the relative concentration $c \to c-c_e$ and similarly $\Psi \to \Psi - \Psi_e$ to get
(using (\ref{genW}) 
\begin{align}
\nonumber
%\label{genWC}
 &\dot{W}_\text{tot} = -\int_0^L (p'(x) \, Q(x) +  \Psi (x)\, Q'(x)) dx \\
 \nonumber
  &= \int_0^L \left(\left( RTc(x)-p(x) \right) Q'(x)  -  Q(x) p'(x)\right)dx\\
   \nonumber
  &= \int_0^L RTc(x) Q'(x)dx - \int_0^L \frac{d}{d x} (pQ) dx \\
\nonumber
% \label{W-tot-pQ}
  &= \int_0^L RTc(x)  Q'(x)dx + \Delta[p\, Q]\\
  \label{W-tot-psi}
  &= \Delta[\Psi\, Q]-\int_0^L RTc'(x) Q(x) dx .
\end{align}
where $\Delta [\Psi\, Q)] = \Psi(0)Q(0)-\Psi(L) Q(L)$ and where we recover (\ref{p-const}) for $c=0$.

If the concentration inside the tube is also constant, the dissipation for the osmotically driven flow is
\begin{align}
\label{W-const-c}
 \dot{W}_\text{tot} &=\Delta[\Psi Q]
\end{align}
and if $Q$ is zero at $x=0$, the analytical solution \cite{Rademaker2017} 
\begin{align}
 Q(x) =&  \frac{2 \pi r_0 L_pL}{m }\,(RTc-p(L))\frac{\sinh \left( m \frac{x}{L} \right)}{\cosh m}\label{eq:qGenX} \\
 RT c -& p(0)=  \frac{ RT c-p(L) }{\cosh m} . \label{eq:pGenX} 
\end{align}
gives the simple form for the dissipation
\begin{align}
\nonumber
\dot{W}_{\text{tot}} &=  - \Psi(L) Q(L) = \left( RTc - p(L)\right) Q(L)\\
 \label{eq:WdotZrMem}
&= 2 \pi r_0 L L_p (R T c -p(L))^2\frac{\tanh m}{m}.
\end{align}
The individual  contributions are similarly
\begin{align}
\nonumber
&\dot{W}_\text{lub}= \\
&  \pi r_0 L L_p (R T c- p(L) )^2\left( -\frac{1}{\cosh^2 m} + \frac{\tanh m}{m}\right) \label{eq:WdotAlt}
\end{align}
and
\begin{align}
\nonumber
&\dot{W}_\text{mem}= \\
&  \pi r_0 L L_p (R T c - p(L) )^2\left(  \frac{1}{\cosh^2 m} + \frac{\tanh m}{m} \right) \label{eq:WdotMem}
\end{align}
so when we add these contributions the $1/\cosh^{2} m$ terms cancel. For small $m$, $\dot{W}_\text{lub}$ is very small ($O(m^2)$)
\begin{align}
\dot{W}_\text{lub} \approx \frac16 m^2 \pi r_0 L L_p (R T c- p(L) )^2 
\end{align}
whereas 
\begin{align}
\dot{W}_\text{mem} \approx \left(2- \frac56m^2\right)\pi r_0 L L_p (R T c- p(L) )^2
\end{align}
so $\dot{W}_\text{mem}$ dominates completely. At large $m$ they become equal:
\begin{align}
\nonumber
\dot{W}_\text{lub} \approx \dot{W}_\text{mem} &\approx \frac{1}{m}\pi r_0 L L_p (R T c- p(L) )^2\\
& = \pi r_0 L_{\text{eff}} L_p (R T c- p(L) )^2 
\end{align}
although $\dot{W}_\text{mem} >\dot{W}_\text{lub} $ for all $m$.

The case $c=0$ corresponds to a porous pipe in a constant external pressure $p_e$. Again, if the pipe is closed at $x=0$ and open at $x=L$ with pressure $p(L) < p_e$, we get
the exact same results, replacing $R T c- p(L)$ by $p_e- p(L)$ in (\ref{eq:WdotZrMem}), (\ref{eq:WdotAlt}) and (\ref{eq:WdotMem}).

\section{Conclusions}

We have studied the viscous energy dissipation in pipe flows with permeable or semipermeable walls in order to generalise the result $\dot W=Q\, \Delta p$ valid for pipes with impermeable walls.
 We have obtained a surprisingly simple expression valid for Stokes flow in long, thin, cylindrical pipe using the slender approximation and representing the porous wall as a collection of cylindrical pores. 
 For a pipe of length $L$, the dissipation given in eqn. (\ref{DissTot}) is expressed in terms of the axially varying flow rate and its derivative as well as the material parameters: pipe radius, wall-permeability and liquid viscosity. 
 For semipermeable pipes, where the water uptake is governed by osmosis, the viscous dissipation, given in  eqn. (\ref{W-tot-psi}), is expressed entirely in terms of the fundamental variables: the
flux, the pressure and the osmotic pressure (or concentration) without any material parameters. This suggests that the result it is much more general than our derivation in terms of cylindrical pores would imply.

\section{acknowledgement}
We are grateful for support from the Danish Council for Independent Research | Natural Sciences (Grant No. 12-126055, {\em Long-distance sugar transport in  trees}) and from Villum Fonden through Research Grant 13166. 

\appendix
\section{Detailed model of the permeability $L_p$ for a cylindrical tube with a porous wall perforated by cylindrical pores.}

\begin{figure}
  \centering
  \includegraphics[]{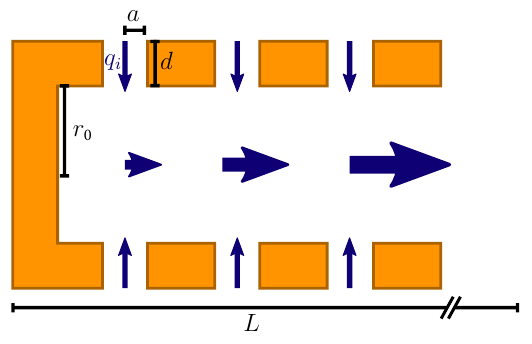} 
  \caption{A sketchl of the flow in the membrane pores showing a tube with pores of radius \(a\) and length \(d\).}
  \label{fig:membrane}
\end{figure}

As an example we can compute the dissipation through a porous tube membrane modelled as a solid surface with \(N\) same-sized, cylindrical pores of radius \(a\) and length $d$, where \(d\) is the thickness of the membrane (see Fig. \ref{fig:membrane}). We expect this model to be useful, even though, in the context of plant leaves the pores (aquaporins) are of nanometric size, which implies that neither the approximation of cylindrical pores nor the validity of the Navier-Stokes equation is well-founded. The density $n$ of pores, per length, is assumed constant, so $n=N/L$. Through each of the pores we assume a Poiseuille flow with resistance
\begin{align}
 R_i = \frac{8\eta d}{\pi a^4}.
\end{align}
The total resistance $R$ of $N$ non-interacting pores in parallel  is related to the permeability \(L_p\) of the membrane as
\begin{align}
 \frac{1}{R} = \frac{N}{R_i} = \frac{N \pi a^4}{8\eta d} \equiv 2\pi r_0 L L_p
\end{align}
giving the relation
\begin{align}
\label{Lp}
L_p =  \frac{n a^4}{16 \eta d r_0}.
\end{align}
or 
\begin{align}
\label{kk}
k= \eta L_p d=  \frac{n a^4}{16 r_0}.
\end{align}
The dissipation inside the pore is dependent on the choice of pore radius \(a\) and covering fraction \(\phi\), since this determines the actual inflow velocity \(v_i\) through pore number $i$ and the corresponding flux $q_i= \pi a^2 v_i$. They are connected to the continuous inflow $v_0(x)$ as 
\be
v_0(x) = \phi \, v_i.
\ee
with the covering fraction 
\begin{align}
\phi = \frac{n \pi a^2}{2\pi r_0 } =  \frac{n  a^2}{2 r_0 } ,
\end{align}
in terms of which 
\begin{align}
L_p  =  \frac{\phi a^2}{4 \eta d } ,
\end{align}
and 
\begin{align}
k  =  \eta L_p d= \frac{1}{4 } \phi a^2,
\end{align}

The viscous dissipation through all pores in the membrane is (by (\ref{dissip2}))
\begin{align}
\dot{W}_\text{mem} &=  \frac{8\eta d}{\pi a^4} \sum_{i=1}^N q_i^2  \notag\\
&= \frac{8\eta d}{\pi a^4} \frac{n \pi^2 a^4}{ \phi^2}\int_0^L v_0^2(x) dx \notag\\
&= \frac{2\pi r_0}{L_p} \int_0^L v_0^2(x) dx \label{eq:contrib2}\\
&= \frac{1}{2\pi r_0 L_p} \int_0^L \left( Q'(x) \right)^2 dx. \label{eq:WdotOfQ}
\end{align}
One might wonder, whether it is valid to retain this term compared to the terms in Eq. (\ref{eq:Wdot03c}), which we discarded.
In particular, the first term in (\ref{eq:Wdot03c}) has precisely the same form as (\ref{eq:WdotOfQ}), but with a different prefactor.
However, $L_p$ is assumed to be small due to the smallness of $a/r_0$ and $\phi$. The ratio of this latter term to (\ref{eq:WdotOfQ})  is  roughly $ (\eta /r_0^2) r_0 L_p =\eta L_p /r_0 =\phi a^2 /(4 r_0)$. The covering fraction $\phi$ must be less than unity (typically it is much less) and since $a/d <1$ and $a/r_0 \ll 1$
this ratio is typically very small.

%\bibliography{bibliography}

\appendix
\end{document}